\documentclass[prb,twocolumn,groupedaddress,floats,citeautoscript,showpacs]{revtex4}

\usepackage{graphicx}
\usepackage{amsmath}
\usepackage{amssymb}
\usepackage{times}
\usepackage{colordvi}

\graphicspath{{.}{./EPS/}}

\newcommand{\ket}[1]{\left |  #1 \right \rangle}

\newcommand{\av}[1]{\langle #1\rangle}

\usepackage{bm}
\newcommand* {\vek}[1]{{\ensuremath{\bm{\mathrm{#1}}}}}
\newcommand* {\ee}{\ensuremath{\mathrm{e}}}

\begin{document}

\title{Anomalous spin-related quantum phase in mesoscopic hole rings}

\author{M. J\"a\"askel\"ainen}
\affiliation{Institute of Fundamental Sciences and MacDiarmid Institute for Advanced
Materials and Nanotechnology, Massey University (Manawatu Campus), Private Bag
11~222, Palmerston North 4442, New Zealand}

\author{U. Z\"ulicke}
\affiliation{Institute of Fundamental Sciences and MacDiarmid Institute for Advanced
Materials and Nanotechnology, Massey University (Manawatu Campus), Private Bag
11~222, Palmerston North 4442, New Zealand}
\affiliation{Centre for Theoretical Chemistry and Physics, Massey University (Albany
Campus), Private Bag 102904, North Shore MSC, Auckland 0745, New Zealand}

\date{\today}

\begin{abstract}

We have obtained numerically exact results for the spin-related geometric quantum
phases that arise in p-type semiconductor ring structures. The interplay between
gate-controllable (Rashba) spin splitting and quantum-confinement-induced mixing
between hole-spin states causes a much higher sensitivity of magnetoconductance
oscillations to external parameters than previously expected. Our results imply a
much-enhanced functionality of hole-ring spin-interference devices and shed new
light on recent experimental findings.

\end{abstract}


\maketitle

\section{Introduction, Motivation \& Summary}

Quantum-interference effects dominate electric transport through conductors that
are {\em mesoscopic\/}, i.e., have a smaller size than the decoherence length
set by inelastic interactions of charge carriers with other degrees of freedom
(e.g., phonons, disorder)~\cite{mesotrans}. In particular, mesoscopic {\em ring\/}
structures exhibit magnetoconductance oscillations~\cite{aboscmetal}
that reveal geometric quantum (Berry~\cite{berry},
Aharonov-Anandan~\cite{ahaanan,anan:nat:92}\-) phases acquired by charge
carriers propagating quantum-coherently through a multiply connected geometry.
Coupling of orbital motion to the spin of charge carriers affects the quantum
interference and geometric phases manifested in charge transport through
rings~\cite{loss:prb:92}. Such spin-dependent electronic
interference effects could form the operational basis for novel transistor
devices~\cite{nitta:apl:99} and quantum logic gates~\cite{foldi:prb:05}.

Strong experimental efforts have been undertaken to identify and measure
spin-related geometric phases in magnetotransport through arrays of mesoscopic
rings~\cite{heida:prl:98,nitta:prl:06}, single ring
structures~\cite{shay:prl:02,yuli:epl:04,mole:prl:06,ensslin:prl:07,habib:apl:07},
and anti-dot superlattices~\cite{katsu:jpsj:07}. Many recent experiments were
performed in {\em p-type\/} semiconductor
structures~\cite{shay:prl:02,ensslin:prl:07,habib:apl:07,katsu:jpsj:07} because
charge carriers from the valence band (holes) are expected to be subject to much
larger momentum-dependent spin splittings than conduction-band
electrons~\cite{rolandbook}. In contrast, many theoretical works have considered 
spin-dependent interference in n-type semiconductor
rings~\cite{frust:prb:04,peet:prb:04,bran:prb:04,vasi:prb:05,tserkov:prb:07,kai:prb:08}, 
while the electronic properties of p-type rings remain largely unexplored. As 
conduction-band electrons and valence-band holes are not merely distinguished by 
the sign of their charge but are known to exhibit very different spin properties, 
especially in quantum-confined structures~\cite{rolandbook}, a careful analysis of 
tunable spin-related quantum phases in p-type mesoscopic rings is needed.

\begin{figure}[b]
\includegraphics[width=3.2in]{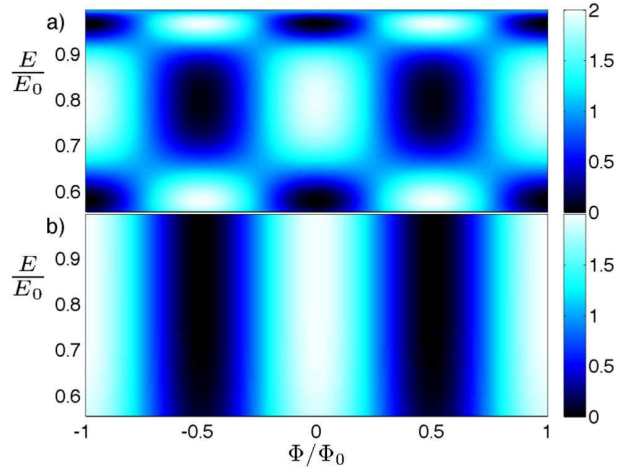}
\caption{\label{fig:magnetoE}
(Color online) Dependence of hole-ring magnetoconductance oscillations on the
Fermi energy $E$. The latter is linearly related to the hole sheet density in the
semiconductor heterostructure and can be controlled by gate voltages. $\Phi$
denotes the magnetic flux penetrating the ring, and $\Phi_0$ is the flux quantum.
Panel~a) shows results from calculations based on our more complete theory,
whereas use of the simplified heavy-hole model for the same ring device yields
panel~b). See text for more details. The range of energies shown corresponds
to the situation where only the lowest ring subband is occupied.}
\end{figure}
Recent calculations~\cite{jairo:prb:07,jairo:prb:08,minchul:prb:09} of the
magnetoconductance in hole rings adopted a purely {\em heavy-hole\/} (HH)
model where only the valence-band states with spin projection $\pm 3/2$ (heavy
holes) and their effective Rashba-type spin splitting are taken into account. It is
tempting to follow such a route because the highest quasi-twodimensional (2D)
valence subband is mostly of HH character for typical hole sheet
densities~\cite{rolandbook}, and the HH model bears resemblance to the one that
applies to conduction-band electrons. However, such an approach neglects the
hole-spin mixing induced by quantum confinement in ring structures. Here we report
results of a theoretical study that fully accounts for spin splitting and spin mixing in
the valence band. Interestingly, we find a synergistic relation between gate-tunable
Rashba spin splitting, which arises from the structural inversion asymmetry (SIA) in
the 2D semiconductor heterostructure, and the hole-spin mixing due to the ring
confinement. This is illustrated in Fig.~\ref{fig:magnetoE} where the dependence of
magnetoconductance oscillations on the Fermi energy in the hole-ring structure is
shown~\cite{note:energyDep}. The more complete theory underlying our calculation
predicts a much more frequent change between maxima and minima of the
magnetoconductance as a function of the Fermi energy than is found for the same
ring geometry and materials parameters within the HH model. Hence, an analysis of
experimental data based on the latter would have to assume an unrealistically large
SIA in the measured sample. Conversely, our results suggest that moderate changes
in hole density and/or SIA, routinely achieved using gate voltages, will be sufficient to
operate a spin-interference-based nanoelectronic device.

This article is organized as follows. Section~\ref{sec:model} presents our
theoretical model for mesoscopic hole rings. Zero-field spin splitting of holes due to
SIA is discussed in Sec.~\ref{sec:spinSplit}. Section~\ref{sec:AAphase} focuses on
the spin-related Aharonov-Anandan phase and how it is revealed in
magneto-conductance oscillations. The frequently
used~\cite{jairo:prb:07,jairo:prb:08,minchul:prb:09} heavy-hole model for
p-type mesoscopic rings is introduced in Sec.~\ref{sec:compHH}, and results
obtained using it are compared with those found within our more complete theory.
The penultimate Sec.~\ref{sec:exp} presents an interpretation of recent experiments
in light of our new results, discussing also possible effects of spin splitting due to
bulk inversion asymmetry (BIA). Our conclusions are given in Sec.~\ref{sec:concl}.
Some relevant mathematical derivations are given in the Appendix.

\section{Model for a mesoscopic hole ring}
\label{sec:model}

\begin{figure}[t]
\includegraphics[width=3.2in]{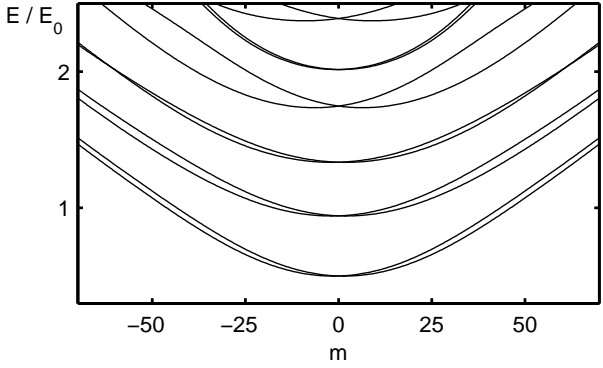}
\caption{\label{fig:subbands}
Hole-ring subbands  in a quantum well without SIA. Parameters: in-plane aspect
ratio (radius/width) $\sqrt{\lambda_R}=20$, $\lambda_d=0.5$ and $\bar\gamma=
0.37$ (value applies to GaAs). $m$ is the eigenvalue of total angular momentum
component $\hat M_z = \hat L_z + \hat J_z$ perpendicular to the ring. $E_0=\pi^2
\hbar^2\gamma_1/(2 m_0 d^2)$ is the energy scale set by size quantization in the
2D quantum well from which the ring is fabricated.}
\end{figure}

Our calculations are based on the $4\times 4$ Luttinger model for the uppermost
valence band in typical semiconductors~\cite{luttham2}, which takes both the
heavy-hole and the light-hole states into account. For simplicity, we neglect band
warping due to the cubic crystal symmetry. The ring confinement is assumed to be
due to a quantum-well potential in $z$ direction (width $d$) and a singular-oscillator
potential $V_\perp(r) = m_0 \omega^2 (r - [R^2/r])^2/2$ for the radial coordinate
$r$ in the $xy$ plane. Here $R$ is the effective ring radius, and the oscillator
potential defines a length scale $\ell_\omega=\sqrt{\sqrt{\gamma_1} \hbar/(m_0
\omega)}$ that is a measure of the in-plane ring width. Only the lowest 2D
quantum-well bound state is taken into account, hence our theory applies in the
(typically realistic) case $d\ll\ell_\omega$. The energy splitting between 2D
heavy-hole and light-hole subband edges is accounted for by the Hamiltonian (we
use the hole picture for the valence band, counting energies as positive from the
bulk valence-band edge)
\begin{equation}
H_{\text{qw}} = \left(1 - 2 \bar\gamma \left[ {\hat J}_z^2 - \frac{5}{4} \right] \right)
E_0 \quad ,
\end{equation}
where $E_0=\pi^2\hbar^2\gamma_1/(2 m_0 d^2)$, $\hat J_z$ is the operator for
the spin-3/2 angular-momentum component perpendicular to the ring plane, and
$\bar\gamma = (2\gamma_2+3\gamma_3)/(5\gamma_1)$ in terms of Luttinger
parameters~\cite{luttham2}. The in-plane hole motion is governed by
\begin{widetext}
\begin{equation}\label{eq:ringHam}
H_{\text{rg}} = \frac{\lambda_d}{4} \left\{ \left(1 + \bar\gamma \left[ {\hat J}_z^2 -
\frac{5}{4} \right] \right) \ell_\omega^2 \hat k_\perp^2 - \bar\gamma \ell_\omega^2
\left( {\hat k}_-^2 {\hat J}_+^2 + {\hat k}_+^2 {\hat J}_-^2 \right) + \left(
\frac{r}{\ell_\omega} - \lambda_R\, \frac{\ell_\omega}{r} \right)^2 \right\} E_0 \quad ,
\end{equation}
\end{widetext}
with $\lambda_d = (2 d/[\pi\ell_\omega])^2$ and $\lambda_R = (R/\ell_\omega)^2$.
Here $\vek{\hat k}_\perp = (\hat k_x, \hat k_y)$ is the in-plane hole wave vector,
${\hat k}_\pm = {\hat k}_x \pm i {\hat k}_y$, and ${\hat J}_\pm = ({\hat J}_x \pm i
{\hat J}_y )/\sqrt{2}$.

The hole-ring Hamiltonian $H_{\text{qw}} + H_{\text{rg}}$ commutes with $\hat M_z
= \hat L_z + \hat J_z$, where $\hat L_z = x\hat k_y - y \hat k_x$. The eigenvalues
$m$ of $\hat M_z$ can thus be used to label states within the quasi-onedimensional
(1D) ring subbands~\cite{uz:prb:08a}. Adopting polar coordinates $r$, $\varphi$ for
the in-plane motion and making the \textit{Ansatz\/} 
\begin{equation}\label{eq:ansatz}
\left\langle r, \varphi \right. \ket{\psi} = \ee^{i (m - \hat J_z) \varphi}\, \chi_{m}(r)
\end{equation}
for the four-spinor hole wave function generates a purely radial Schr\"odinger
equation that we solve numerically using a pseudospectral
method~\cite{pseudoSpectral,weideman:2000:MDM} tailored to our needs.
Figure~\ref{fig:subbands} shows a representative result for ring subbands
$E^{(n)}_s(m)$, where $s=\pm 1$ distinguishes spin-split dispersions with
eigenvalues related via $E^{(n)}_s(m)=E^{(n)}_{-s}(-m)$, and $n=0, 1,
\dots$ labels the doublets starting with the lowest-lying one.

\section{Effects of SIA spin splitting}
\label{sec:spinSplit}

To investigate spin-related geometric phases in hole rings, we include the
dominant SIA contribution to the bulk-hole Hamiltonian~\cite{rolandbook}, which
is given by ${\mathcal H}_{\text{SIA}}^{(\text{bulk})} = r_{41}^{8v8v} \left(
\vek{\hat k} \times \vek{\mathcal{E}} \right)\cdot \vek{\hat J}$. In our case of
interest, the SIA electric field $\vek{\mathcal{E}}$ has a $z$ component
${\mathcal E}_z$ determined by the 2D quantum-well confinement. In addition,
the radial in-plane (ring) confinement induces an SIA spin splitting. We find
${\mathcal H}_{\text{SIA}}^{(\text{rg})} = H_{\text{SIA}}^{(\text{rg,qw})} +
H_{\text{SIA}}^{(\text{rg,rg})}$, with
\begin{subequations}
\begin{eqnarray}
H_{\text{SIA}}^{(\text{rg,qw})} &=& \frac{V_{\text{SIA}}}{V_{41}} \,
\sqrt{\frac{\lambda_d}{2}} \, \ell_\omega \, i \left( \hat k_+ \hat J_- - \hat k_-
\hat J_+ \right) E_0 \,\, , \\
\label{ringSIArg}
H_{\text{SIA}}^{(\text{rg,rg})} &=& \lambda_{41} \, \frac{\lambda_d}{2} \left[
1 - \lambda_R^2 \left(\frac{\ell_\omega}{r}\right)^4\right] \hat L_z \, \hat J_z \, E_0
\,\, .
\end{eqnarray}
\end{subequations}
The voltage $V_{\text{SIA}}={\mathcal E}_z d$ is a measure for SIA in the quantum
well, and $V_{41}=\pi \gamma_1 \hbar^2 / (m_0 |r_{41}^{8v8v}|)$ is a materials
parameter (11.44~V in GaAs). $\lambda_{41}=|r_{41}^{8v8v}|/(e\ell_\omega^2)$ is
typically very small unless the ring becomes narrow on the scale
$\sqrt{|r_{41}^{8v8v}|/e}$, which is of order \AA. We checked that matrix elements
of (\ref{ringSIArg}) are negligible for typical hole-ring device parameters. Thus, SIA
splitting due to the in-plane ring confinement can be disregarded. 

The hole-ring Hamiltonian including the SIA terms still commutes with $\hat M_z$.
Using the same procedure as outlined in Sec.~\ref{sec:model}, we obtain the
ring-subband dispersions including the SIA term $H_{\text{SIA}}^{(\text{rg,qw})}$.

\section{Aharonov-Anandan phase and magneto-conductance oscillations}
\label{sec:AAphase}

\begin{figure}[t]
\includegraphics[width=3.2in]{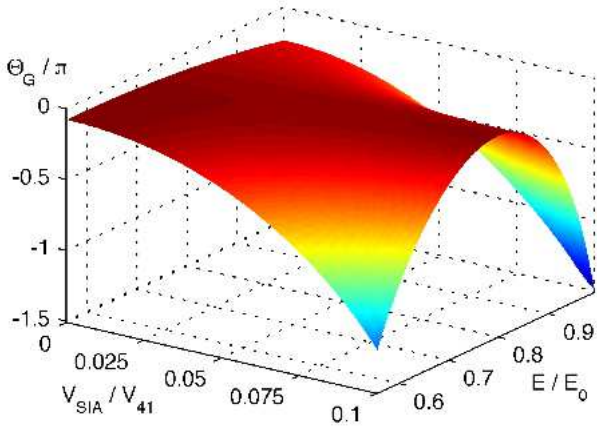}
\caption{\label{fig:phase}
(Color online) Spin-related quantum (Aharonov-Anandan) phase for holes in a
mesoscopic ring, plotted as a function of hole energy $E$ and the voltage
$V_{\text{SIA}}$ associated with SIA in the semiconductor heterostructure
from which the ring is fabricated. Other parameters are the same as in
Fig.~\ref{fig:subbands}. The energy range shown corresponds to the situation
where only the lowest ring subband is populated. The band-structure
parameter $V_{41}$ is 11.44~V in GaAs.}
\end{figure}

Knowledge of the ring-subband dispersions makes it possible~\cite{nitta:apl:99,
frust:prb:04,peet:prb:04,jairo:prb:07,uz:prb:08a} to extract the Aharonov-Anandan
(AA) phase~\cite{ahaanan} for holes traversing the ring at a particular energy $E$
(generally, the Fermi energy of holes in the 2D semiconductor heterostructure).
In the following, we focus entirely on the energy range where only the lowest
spin-split ring subband is relevant, but our results can easily be generalized.
In terms of the two Fermi angular momenta $m_\pm(E)$ defined by
$E=E_s^{(0)}(\pm s \, m_\pm)$, the AA phase is given by (see the Appendix for
details of the derivation)
\begin{equation}\label{eq:AAphase}
\Theta_{\text{G}} = \pi \left( m_+ - m_-  - 3 \right) \quad .
\end{equation}
The dependence of this spin-related geometric phase on the Fermi energy $E$
and SIA strength $V_{\text{SIA}}$ is shown in Fig.~\ref{fig:phase} for a set of
typical hole-ring parameters.

\begin{figure}[t]
\includegraphics[width=3.2in]{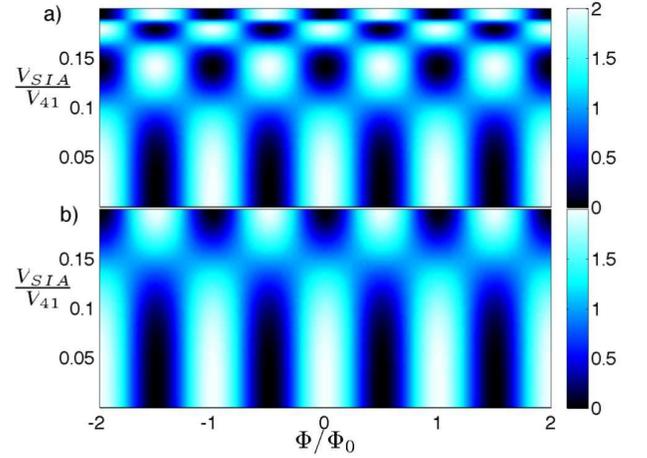}
\caption{\label{fig:magnetoSIA}
(Color online) Magnetoconductance oscillations for a mesoscopic hole ring
attached to ideal leads, shown as a function of the voltage $V_{\text{SIA}}$
associated with SIA in the semiconductor heterostructure. Panel~a) is obtained
from our more complete theory, whereas panel~b) results from application of
the HH model to the same ring device. The band-structure parameter $V_{41}$
is 11.44~V in GaAs. In the calculation, the Fermi level was assumed to be at
$0.6 E_0$, and the visibility $A$ was set to $1$. Other parameters are as for
Fig.~\ref{fig:subbands}.}
\end{figure}
At low-enough temperatures, the electric conductance $G$ of a mesoscopic
ring attached to ideal leads~\cite{note:idLead} exhibits a quantum-interference
contribution that makes it possible to measure the Aharonov-Anandan phase.
Quite generally, it is given by the expression (see the Appendix for more details)
\begin{equation}\label{eq:MagCond}
G = G_0 \left[ 1 + A \cos \left( 2\pi \frac{\Phi}{\Phi_0} \right) \cos \Theta_{\text{G}}
\right ] \quad ,
\end{equation}
where $A\le 1$ measures the visibility of quantum interference in the ring. The
first cosine term contains the magnetic flux $\Phi$ penetrating the ring's area,
measured in units of the flux quantum $\Phi_0=2\pi\hbar/e$. It gives rise to
magnetoconductance oscillations~\cite{gefen:prl:84}  that are the
electric analogue of the Aharonov-Bohm effect~\cite{ahabohm}. The modulation
of the magnetoconductance as a function of SIA strength achieved, e.g., by
external gate voltages, reveals the presence of the spin-related quantum phase
$\Theta_{\text{G}}$. This is illustrated in Fig.~\ref{fig:magnetoSIA}a.

Experimentally, a change in the strength of SIA while keeping all other
parameters (in particular, the hole density) constant can be achieved by
simultaneously applied front and back-gate voltages~\cite{grund:prl:00}. However,
in the majority of samples~\cite{shay:prl:02,nitta:prl:06,mole:prl:06,habib:apl:07},
only a single (front or back) gate is available. In such a situation, both SIA and
the density of charge carriers in the semiconductor heterostructure are changed
by a gate voltage. For holes, changing the density (i.e., the Fermi energy) has a
profound effect on the spin-related quantum phase. This can be inferred from the
strong dependence of $\Theta_{\text{G}}$ on $E$ for constant $V_{\text{SIA}}$
seen in Fig.~\ref{fig:phase}. The modulation of magnetoconductance oscillations
when changing {\em only\/} the hole density (keeping $V_{\text{SIA}}=0.1\, V_{41}$
constant, and with $A=1$) is illustrated in Fig.~\ref{fig:magnetoE}a. Upto a constant
shift, $E$ is directly proportional to the 2D sheet density of holes in the heterostructure.

\section{Comparison with the heavy-hole model}
\label{sec:compHH}

The necessity to fully account for valence-band mixing in hole rings can be
illustrated by a direct comparison with the simpler HH model. The latter
results from a perturbative (L\"owdin-partitioning) treatment~\cite{rolandbook}
of valence-band mixing and SIA splitting for the lowest 2D HH subband,
neglecting further spin splitting and mixing due to the in-plane confinement.
The Hamiltonian of the HH-model ring is $H_{\text{qw}}^{\text{(HH)}}
+ H^{\text{(HH)}}_{\text{rg}} + H^{\text{(HH)}}_{\text{SIA}}$, with
\begin{subequations}
\begin{eqnarray}
H_{\text{qw}}^{\text{(HH)}} &=& \left( 1 - 2 \bar\gamma \right) E_0 \quad , \\
H^{\text{(HH)}}_{\text{rg}} &=& \frac{\lambda_d}{4} \left\{ \left( 1 +  \bar
\gamma \right) \ell_\omega^2 \, {\hat k}_\perp^2 + \left( \frac{r}{\ell_\omega} -
\lambda_R\, \frac{\ell_\omega}{r} \right)^2 \right\} E_0 \,\, , \nonumber \\ \\
H^{\text{(HH)}}_{\text{SIA}} &=& \frac{V_{\text{SIA}}}{V_{41}} \, \left(
\frac{\lambda_d}{8}\right)^{\frac{3}{2}} \ell_\omega^3 \,\, i \left( \hat k_+^3 \hat
J_-^3 - \hat k_-^3 \hat J_+^3 \right) E_0 \,\, , \\
& \equiv & \frac{3}{4}\, \frac{V_{\text{SIA}}}{V_{41}} \left( \frac{\lambda_d}{4}
\right)^{\frac{3}{2}} \ell_\omega^3  \,\, i \left( \hat k_+^3 \hat \sigma_- - \hat k_-^3
\hat \sigma_+ \right) E_0 \,\, . 
\end{eqnarray}
\end{subequations}
The second line defining $H^{\text{(HH)}}_{\text{SIA}}$ applies when the HH
($J_z = \pm 3/2$) amplitudes are treated as an effective spin-1/2 degree of
freedom; this is the way SIA spin splitting for heavy holes is usually
written~\cite{rolandbook}. No coupling to light-hole amplitudes is present in
the HH model, even after the in-plane ring confinement is introduced.

Using the same numerical method as for the full spin-3/2 Luttinger theory of
hole rings outlined above, we find the subbands of the HH-model ring and the
spin-related quantum phase associated with the lowest one. Its dependence on
both energy $E$ and strength of SIA turns out to be much weaker than in the more
complete theory. For the energy dependence of magnetoconductance oscillations,
this is illustrated in Fig.~\ref{fig:magnetoE} for a ring with $V_{\text{SIA}}=0.1\,
V_{\text{41}}$ and all other parameters as in Fig.~\ref{fig:subbands}. A similar
result is obtained when energy $E$ is fixed and $V_{\text{SIA}}$ is varied. See
Fig.~\ref{fig:magnetoSIA}. The different behavior exhibited by the full Luttinger
model as compared with the HH model arises from HH-LH mixing induced by
the in-plane ring confinement. Hence, differences in quantitative predictions
from the two models scale with $\lambda_d$ and thus vanish in the 2D limit.

\section{Application to real hole-ring samples}
\label{sec:exp}

\begin{table}
\caption{\label{tab:exp}
Phase shift of magnetoconductance oscillations induced by varying
$V_{\text{SIA}}/V_{41}$ between 0.527 and 0.551, as measured for a 1D GaAs hole
ring (Ref.~~\onlinecite{habib:apl:07}) and calculated using the full Luttinger model
and the simpler HH model, respectively. Parameters used in the calculations are
$\bar\gamma = 0.37$, $\lambda_d = 1.0$, $\lambda_R = 28$, and $E=1.1\, E_0$
(Luttinger model), $n_{\text{F}}=1$ (HH model). In general, $E$ can be determined
from the 2D hole sheet density and details of the sample's quantum-well confinement.}
\begin{ruledtabular}
\begin{tabular}{ccc}
 experiment & \mbox{Luttinger model} & \mbox{HH model} \\ \hline
  $\pi$	&  $0.25\,\pi$ 	& $0.13\,\pi$
\end{tabular}
\end{ruledtabular}
\end{table}

Our theory enables a more detailed quantitative interpretation of experimental
results. Applied gate voltages have been observed to shift magnetoconductance
oscillations~\cite{mole:prl:06,habib:apl:07,moleExpNote}. Comparison with
Shubnikov-de~Haas data measured in the unstructured 2D HH system
enabled experimentalists to quantify the change in SIA strength required for
a $\pi$ phase shift. The HH model predicts $\Delta V_{\text{SIA}}=V_{41} /
(n_{\text{F}}^2\lambda_d\lambda_R)^{1/2}$ for a ring with $n_{\text{F}}$ occupied
1D subbands~\cite{jairo:prb:07,jairo:prb:08}. Interestingly, the experiment reported in
Ref.~~\onlinecite{habib:apl:07} observed an order-of-magnitude discrepancy between 
the measured value and that expected from application of the HH model. As the 
comparison given in Table~\ref{tab:exp} shows, taking into account the enhancement
due to HH-LH mixing within the full Luttinger model markedly reduces this
discrepancy. We suspect that even better agreement could be reached if (a)~more
details about the ring structure were known, thus facilitating a more realistic modeling
of the quantum-well and in-plane confinement potentials, and (b)~the effect of
band-warping corrections were included. Finally, typical ring devices are fabricated in
semiconductors whose unit cell lacks inversion symmetry and, thus, are subject to an additional spin splitting due to BIA. We will briefly discuss BIA effects before concluding.

The most important BIA spin-splitting term in the bulk-hole Hamiltonian is~\cite{rolandbook} 
${\mathcal H}^{\text{(bulk)}}_{\text{BIA}} = b_{41}^{8v8v} \left( \left\{ k_x, k_y^2 - k_z^2
\right\} J_x + \text{c.p.} \right)$. Introducing the 2D quantum-well confinement by replacing
$\hat k_z \to \av{\hat k_z}\equiv 0$ and $\hat k_z^2 \to \av{\hat k_z^2}\equiv \pi^2/d^2$
yields the BIA contribution to the model-ring Hamiltonian as $H_{\text{BIA}}^{\text{(rg)}}= H_{\text{BIA}}^{\text{(rg,qw)}}+H_{\text{BIA}}^{\text{(rg,rg)}}$, where
\begin{subequations}
\begin{eqnarray}
H_{\text{BIA}}^{\text{(rg,qw)}} &=&  \frac{\ell_{\text{BIA}}}{d} \, \sqrt{\frac{\lambda_d}{2}}
\,\, \ell_\omega \left( \hat k_+ \hat J_+ + \hat k_- \hat J_- \right) \quad ,\\
H_{\text{BIA}}^{\text{(rg,rg)}} &=&   \frac{\ell_{\text{BIA}}}{d} \left( \frac{\lambda_d}{8}
\right)^{\frac{3}{2}} \ell_\omega^3 \left( \hat k_+^2 - \hat k_-^2 \right) \left(\hat k_+
\hat J_- - \hat k_- \hat J_+ \right) . \nonumber \\
\end{eqnarray}
\end{subequations}
The length scale $\ell_{\text{BIA}}=\pi m_0 |b_{41}^{8v8v}|/(\gamma_1\hbar^2)\equiv
4.83$~\AA\ in GaAs. As typical quantum-well widths are of the order of $10$~nm, we
have $\ell_{\text{BIA}}/d \sim 0.05$. This value is an order of magnitude smaller than
$V_{\text{SIA}}/V_{41}$ measured in GaAs ring samples with the strongest Rashba
splitting~\cite{habib:apl:07}. Hence, as a first approximation, it is admissible to neglect
BIA spin splitting when discussing this experiment.

Formally, the BIA terms do not commute with $\hat M_z$, and the \textit{Ansatz\/}
given in Eq.~(\ref{eq:ansatz}) will not eliminate the $\varphi$-dependence from the
BIA part of the ring Hamiltonian. In essence, previous eigenstates with quantum
number $m$ are coupled via the BIA term to those with $m \pm 2$. A reduced-band
model may be adequate to explore BIA effects in the lowest ring subband.

\section{Conclusions}
\label{sec:concl}

We have obtained numerically exact results for electronic subbands and spin-related
geometric phases for holes in mesoscopic rings. Unlike previous models, we account
fully for spin splitting and mixing arising in the quantum-confined valence band. For
quasi-onedimensional ring structures, a much stronger modulation of magnetoconductance
oscillations (as a function of Fermi energy and/or SIA spin-splitting strength) is found as
compared with simplified (purely HH) models. This effect arises due to HH-LH mixing
induced by the in-plane ring confinement, and the magnitude of the enhanced dependence
is quantified by the parameter $\lambda_d$, which is related to the ratio of quantum-well
width and in-plane ring width.

We have applied our model to discuss a recent experiment where an anomalously
strong modulation of Aharonov-Bohm oscillations was observed. A sizable
enhancement of magnetoconductance-oscillation modulations is obtained within our (on
some level still idealized) model, but its magnitude is smaller than the observed value.
A more realistic modelling of the ring structure may be needed to reach full agreement.
We also ascertained the effect of BIA spin splitting. The parameter quantifying its
importance is the ratio of a length scale $\ell_{\text{BIA}}$ ($=4.83$~\AA\ in GaAs) and
the quantum-well width, which was negligible compared to the strength of SIA splitting
present in the experiment under consideration. Our theory, possibly with further
refinement, should be useful for guiding efforts~\cite{nitta:apl:99,foldi:prb:05} aimed at
realizing novel electronic devices based on spin-dependent quantum interference.

\begin{acknowledgments}
This work is supported by the Marsden Fund Council (contract MAU0702) from 
Government funding, administered by the Royal Society of New Zealand.
\end{acknowledgments}

\appendix*
\section{Derivation of expressions given for the Aharonov-Ananadan phase and
the magnetoconductance}

We assume a standard two-terminal transport geometry as shown, e.g.,
in Fig.~2 of Ref.~~\onlinecite{uz:prb:08a}. To keep the notation uncluttered,
we consider the situation where only the lowest spin-split subband is relevant,
but all formulae can be straighforwardly generalized to the multi-subband case.
Holes with energy $E$ are injected by an external lead at $\varphi=0$ in channel
$s$ in a superposition of ring-state amplitudes: $\ket{\text{in}}_s =
\xi_{s+}^{\text{(in)}}\chi_{s m_+}^{(s)}+\xi_{s-}^{\text{(in)}}\chi_{-s m_-}^{(s)}$.
Here $\chi_m^{(s)}$ denotes the radial four-spinor wave function [see
Eq.~(\ref{eq:ansatz})] for an eigenstate from subband $s$, and the coefficients
$\xi_{s\pm}^{\text{(in)}}$ depend on details of the coupling between ring and
injecting lead. At a draining lead located diametrically opposite to the injecting
one, holes from channel $s$ will enter with an amplitude
\begin{eqnarray}
\ket{\text{out}}_s &=& \ee^{i \pi (m_+ - s\hat J_z)} \, \xi_{s+}^{\text{(in)}}
\chi_{s m_+}^{(s)} +\ee^{i \pi (m_- + s\hat J_z)} \, \xi_{s-}^{\text{(in)}}
\chi_{-s m_-}^{(s)} \nonumber \\
&\equiv& s i {\mathcal M} \left( \ee^{i\pi m_+} \, \xi_{s+}^{\text{(in)}}
\chi_{s m_+}^{(s)} - \ee^{i\pi m_-} \, \xi_{s-}^{\text{(in)}} \chi_{-s m_-}^{(s)} \right) ,
\nonumber \\ \end{eqnarray}
with the matrix ${\mathcal M}=\text{diag}\{1, -1, 1, -1\}$. Thus the phase difference
between forward- and backward-propagating amplitudes is found to be
$s\Theta_{\text{G}}$, with $\Theta_{\text{G}}$ given by Eq.~(\ref{eq:AAphase}).
We have used the freedom that phases are determined only modulo integer
multiples of $2\pi$ to adjust $\Theta_{\text{G}}$ such that it vanishes in the limit
where $V_{\text{SIA}}=0$ and HH-LH mixing is neglected. 

When the ring is penetrated by a magnetic flux $\Phi$ and coupled to ideal leads,
the probability for transmission of holes is obtained as~\cite{uz:prb:08a,magFieldCav}
\begin{equation}
T=\left| \sum_s \left[ \ee^{i \pi \left( m_+ + s \frac{\Phi}{\Phi_0}\right)}\,
\xi_{s+}^{\text{(in)}}\, \xi_{s+}^{\text{(out)}} - \ee^{i \pi \left( m_- - s \frac{\Phi}{\Phi_0}
\right)}\, \xi_{s-}^{\text{(in)}} \, \xi_{s-}^{\text{(out)}}\right] \right|^2  ,
\end{equation}
where the factors $\xi_{s\pm}^{\text{(out)}}$ depend on the coupling between
states $s {\mathcal M} \chi_{s m_\pm}^{(s)}$ and the scattering state in the outgoing
lead. The two-terminal ring conductance $G_{\text{rg}} = \frac{e^2}{2\pi\hbar} T$ is
then given, in full generality, by
\begin{eqnarray}
G_{\text{rg}} &=& G_0 \left[ 1  + \sum_s A_s^{(1)} \cos \left( 2\pi \frac{\Phi}{\Phi_0}
+ s \Theta_{\text{G}} \right) \right. \nonumber  \\
&& \hspace{1cm} \left. + A^{(2)} \cos \Theta_{\text{G}} + A^{(3)} \cos \left( 2\pi
\frac{\Phi}{\Phi_0} \right)  \right] . \qquad
\end{eqnarray}
The familiar contributions proportional to $A_s^{(1)}$ arise from interference between
counter-propagating amplitudes from the same channel and manifest the AA phase.
Additional interference terms (those proportional to $A^{(2,3)}$) are possible because
coupling to leads may induce a mixing conductance between the two channels. In
practice, this happens when the Hilbert space spanned by scattering states in the leads
does not fully contain the space spanned by ring eigenstates. Such a situation could
occur, in principle, because of the sensitive dependence of hole states on quantum
confinement in the leads. As such effects will be small in typical situations and also
depend strongly on the particular realizations of ring-lead couplings, we have not
considered them further in the context of this work. For similar reasons, we assume
that the leads couple symmetrically to states in the two subbands. Setting $A^{(2)}=
A^{(3)} =0$ and $A_+^{(1)}=A_{-}^{(1)}=A/2$ yields Eq.~(\ref{eq:MagCond}), after
application of an addition theorem for cosine functions.


\begin{thebibliography}{35}
\bibitem{note:energyDep}
Spin-dependent geometric phases due to SIA are generally energy-dependent in
hole systems -- a distinguishing feature not shared by conduction-band electrons.
\bibitem{moleExpNote}
In principle, our theory applies to the n-type device of Ref.~~\onlinecite{mole:prl:06}
because a band inversion in the heterostructure from which the ring was fabricated
renders the conduction band to have valence-band-like properties. However, this
structure appears to be quite close to the 2D limit so that HH-LH mixing may not be
very important.
\bibitem{magFieldCav}
In principle, all orbital magnetic-field effects and Zeeman splitting are accounted
for by our expressions if the $m_\pm$ are obtained from ring subbands that
have been calculated with all pertinent terms in the Hamiltonian included. For
simplicity, we neglect Zeeman splitting and magnetic-field-related distortions
of in-plane bound-state wave functions in this work. Our results are directly
applicable to rings with sufficiently large aspect ratio where magnetoconductance
oscillations can already appear in the weak-field regime.
\end{thebibliography}

\end{document}